\documentclass{jors}

\definecolor{mygray}{gray}{0.6}

\usepackage{amsmath}
\usepackage[numbers]{natbib}
\usepackage{amsfonts}
\usepackage{graphicx}
\usepackage{placeins}
\usepackage{booktabs}
\usepackage{listings}

\makeatletter
\newcommand{\gobblecomma}{\@ifnextchar,{\@gobble}{}}
\makeatother

\begin{document}

{\bf Software paper for submission to the Journal of Open Research Software} \\

\section*{(1) Overview}

\vspace{0.5cm}

\section*{Title}

PDSim: A Shiny App for Simulating and Estimating Polynomial Diffusion Models in Commodity Futures

\section*{Paper Authors}

1. He, Peilun; (corresponding author) \\
2. Kordzakhia, Nino; \\
3. Peters, Gareth W.; \\
4. Shevchenko, Pavel V.

\section*{Paper Author Roles and Affiliations}
1. PhD, Department of Actuarial Studies and Business Analytics, Macquarie University, Australia \\
2. Senior Lecturer, School of Mathematical and Physical Sciences, Macquarie University, Australia \\
3. Professor, Department of Statistics and Applied Probability, University of California Santa Barbara, United States \\
4. Professor, Department of Actuarial Studies and Business Analytics, Macquarie University, Australia

\section*{Abstract}

PDSim is an R package that enables users to simulate commodity futures prices using the polynomial diffusion model introduced in {Filipovi{\'c} and Larsson \cite{filipovic2016polynomial} through both a Shiny web application and R scripts. For user-supplied data, a standalone R routine has been developed to provide joint estimation of state variables and model parameters via the Extended Kalman Filter (EKF) or Unscented Kalman Filter (UKF). With its user-friendly interface, PDSim makes the features of simulations and estimations accessible. To date, it is the only package specifically designed for the simulation and estimation of the polynomial diffusion model. The Schwartz-Smith two-factor model \cite{schwartz2000short} is also available within this package for both simulation and calibration.

The package is validated through several tests, including replication of the results in \cite{schwartz2000short}, unit testing of the coverage rate, and verification of the outputs of the main functions.}

\section*{Keywords}

Polynomial diffusion model; Schwartz-Smith model; commodity futures; two-factor commodity modelling; financial simulation

\section*{Introduction}

Stochastic models are crucial in the analysis of commodity futures, serving important roles in various financial areas such as price forecasting, risk management, and portfolio optimisation. Typically, the underlying spot price, denoted as $S_t$, is modelled as a function of certain factors. In 1990, the Ornstein-Uhlenbeck (OU) process was introduced to model oil futures in a two-factor setup, representing spot price and convenience yield \cite{gibson1990stochastic}. Building on this foundation, Schwartz and Smith \cite{schwartz2000short} modelled the logarithm of the underlying spot price of crude oil futures as the sum of two latent factors. These factors, assumed to follow the OU process, capture short-term fluctuations and the long-term equilibrium price level, respectively. This latent factor model and its extensions have been widely used in stochastic modelling in various fields, including biology \cite{favetto2010parameter}, electricity forwards \cite{kiesel2009two}, agricultural commodity futures \cite{sorensen2002modeling}, and crude oil futures \cite{ames2020risk, cortazar2006n, peters2013calibration}.

In this framework, the positivity of the futures price is ensured via an exponential transformation. However, in some markets, prices can become negative. For instance, in 2019, the natural gas price in the Permian Basin, West Texas, turned negative several times. On 20 April 2020, the front-month May 2020 WTI crude oil futures settled at –\$37.63 per barrel on the New York Mercantile Exchange, see, e.g., \cite{fernandez2023negative}. Negative prices occur even more frequently in the electricity market. In 2009, the province of Ontario, Canada, recorded 280 hours of negative electricity prices. In 2017, German electricity prices turned negative over 100 times. Between early 2018 and late 2020, electricity prices in Texas, USA, were negative for nearly 19\% of total hours. Such cases illustrate that the previously mentioned framework is not applicable to these markets.

The polynomial diffusion model relaxes this assumption. In this framework, the spot price is represented as a polynomial of any order in terms of the factors. In particular, under certain conditions, it can be proven that the conditional expectation of the spot price, which is equivalent to the futures price under the assumption of an arbitrage-free market and non-stochastic interest rate, is also a polynomial in terms of factors. The mathematical foundations of the polynomial diffusion model were introduced in \cite{filipovic2016polynomial}, with applications of this framework seen in the modelling of electricity forwards, where the spot price is represented by a quadratic form of two factors \cite{kleisinger2020multifactor}.

In this paper, we introduce the R package PDSim, which is specifically designed to simulate futures prices and estimate latent factors using the polynomial diffusion model. PDSim consists of two components: the core package and a Shiny application of the same name. The core package provides exported functions for simulation as well as for linear and non-linear filtering. It is intended for users who wish to simulate data or fit real-world data. The Shiny app, by contrast, is designed exclusively for simulating data through a graphical user interface (GUI) and for visualising the results in a straightforward manner. This app has also been deployed on a Shiny server, allowing users to access it without requiring any local installation. This separation reflects the computational demands of parameter estimation when working with real-world data. Given the high dimensionality of the parameter space, estimation is computationally expensive, making R scripts a more suitable choice than a web-based interface in such cases.

In this paper, we provide a guide for using the GUI to simulate data and examples of how to use R scripts to both simulate and fit real-world data. The primary focus of PDSim is on simulation and joint estimation of state variables and model parameters, enabling users to explore the properties of the polynomial diffusion model. For applications involving real-world data, we recommend combining PDSim with established optimisation algorithms to minimise the negative log-likelihood returned by the KF, EKF, or UKF functions.

\section*{I. Schwartz-Smith two-factor model}

Under the Schwartz-Smith framework, the logarithm of spot price $S_t$ is modelled as the sum of two factors $\chi_t$ and $\xi_t$, 
\begin{equation}\label{eq:st}
\log{(S_t)} = \chi_t + \xi_t,
\end{equation}
where $\chi_t$ represents the short-term fluctuation and $\xi_t$ is the long-term equilibrium price level. Additionally, we assume both $\chi_t$ and $\xi_t$ follow a risk-neutral Ornstein-Uhlenbeck (OU) process, 
\begin{equation}
d\chi_t = (-\kappa \chi_t - \lambda_{\chi}) dt + \sigma_{\chi} d W_t^{\chi *}, 
\label{eq:SS_rn_chi}
\end{equation}
and 
\begin{equation}
d\xi_t = (\mu_{\xi} - \gamma \xi_t - \lambda_{\xi})dt + \sigma_{\xi} dW_t^{\xi *},
\label{eq:SS_rn_xi}
\end{equation}
where $\kappa, \gamma \in \mathbb{R}^+$ are the speed of mean-reversion parameters, $\mu_{\xi} \in \mathbb{R}$ is the mean level of the long-term factor, $\sigma_{\chi}, \sigma_{\xi} \in \mathbb{R}^+$ are the volatility parameters, and $\lambda_{\chi}, \lambda_{\xi} \in \mathbb{R}$ are risk premiums. The processes $(W_t^{\chi *})_{t \ge 0}$ and $(W_t^{\xi *})_{t \ge 0}$ are correlated standard Brownian Motions with 
$$\mathbb{E} \left(dW_t^{\chi *} dW_t^{\xi *}\right) = \rho dt. $$
In the original Schwartz-Smith model \cite{schwartz2000short}, $\gamma = 0$, and only the short-term factor $\chi_t$ follows an OU process. Since 2000, many studies such as \cite{ames2020risk, casassus2005stochastic, manoliu2002energy, peters2013calibration} and others considered the extended version of the Schwartz-Smith model by introducing $\gamma \ge 0$. For simplicity, further, we continue to refer to this model with $\gamma \ge 0$ as the Schwartz-Smith model. Setting $\lambda_{\chi} = \lambda_{\xi} = 0$ in \autoref{eq:SS_rn_chi} and \autoref{eq:SS_rn_xi} gives the ``real processes'' of the factors under statistical measure. The risk-neutral processes are used to calculate futures prices, and the real processes are for modelling state variables in real time. 

In discrete time, $\chi_t$ and $\xi_t$ are jointly normally distributed. Therefore, the spot price is log-normally distributed. Moreover, under the arbitrage-free assumption and non-stochastic interest rate, the futures price $F_{t,T}$ at current time $t$ with maturity time $T$ must be equal to the expected value of spot price at maturity time $T$, 
$$F_{t,T} = \mathbb{E}^*(S_T | \mathcal{F}_t), $$
where $\mathcal{F}_t$ is the information known at time $t$ and $\mathbb{E}^*(\cdot)$ is the expectation under the risk-neutral processes from \autoref{eq:SS_rn_chi} and \autoref{eq:SS_rn_xi}. Then we can get the linear Gaussian state space model: 
\begin{equation}
x_t = c + Ex_{t-1} + w_t, 
\end{equation}
\begin{equation}
y_t = d_t + F_t x_t + v_t, 
\end{equation}
where 
$$x_t = \left[ \begin{matrix} \chi_t \\ \xi_t \end{matrix} \right], c = \left[ \begin{matrix} 0 \\ \frac{\mu_{\xi}}{\gamma} \left(1 - e^{-\gamma \Delta t} \right) \end{matrix} \right], E = \left[ \begin{matrix} e^{-\kappa \Delta t} & 0 \\ 0 & e^{-\gamma \Delta t}\end{matrix} \right],$$ 
$$y_t = \left( \log{(F_{t,T_1})}, \dots, \log{(F_{t,T_m})} \right)^\top, d_t = \left( A(T_1 - t), \dots, A(T_m - t) \right)^\top,$$
$$F_t = \left[ \begin{matrix} e^{-\kappa (T_1 - t)}, \dots, e^{-\kappa (T_m - t)} \\ e^{-\gamma (T_1 - t)}, \dots, e^{-\gamma (T_m - t)} \end{matrix} \right]^\top$$
and $m$ is the number of futures contracts. The function $A(\cdot)$ is given by 
\begin{align}
A(\tau) =& -\frac{\lambda_{\chi}}{\kappa}(1 - e^{-\kappa \tau}) + \frac{\mu_{\xi} - \lambda_{\xi}}{\gamma}(1 - e^{-\gamma \tau}) \nonumber \\
&+ \frac{1}{2}\left(\frac{1 - e^{-2\kappa \tau}}{2\kappa}\sigma_{\chi}^2 + \frac{1 - e^{-2\gamma \tau}}{2\gamma}\sigma_{\xi}^2 + 2\frac{1 - e^{-(\kappa + \gamma)\tau}}{\kappa + \gamma}\sigma_{\chi}\sigma_{\xi}\rho \right), \nonumber
\end{align}
where $\tau$ is the time to maturity; $w_t$ and $v_t$ are multivariate Gaussian noise sequences with mean $\textbf{0}$ and covariance matrix $\Sigma_w$ and $\Sigma_v$ respectively, where 
$$\Sigma_w = \left[ \begin{matrix}
\frac{1 - e^{-2\kappa \Delta t}}{2\kappa}\sigma_{\chi}^2 & \frac{1 - e^{-(\kappa + \gamma) \Delta t}}{\kappa + \gamma}\sigma_{\chi}\sigma_{\xi}\rho \\
\frac{1 - e^{-(\kappa + \gamma) \Delta t}}{\kappa + \gamma}\sigma_{\chi}\sigma_{\xi}\rho & \frac{1 - e^{-2\gamma \Delta t}}{2\gamma}\sigma_{\xi}^2
\end{matrix} \right],$$ 
and we assume $\Sigma_v$ is diagonal, i.e., 
$$
\Sigma_v = \left[ \begin{matrix}
\sigma_1^2 & 0 & \dots & 0\\
0 & \sigma_2^2 & \dots & 0 \\
\vdots & \vdots & \ddots & \vdots \\
0 & 0 & \dots & \sigma_m^2
\end{matrix} \right].
$$
$\Delta t$ is the time step. Under this framework, $c, E, \Sigma_w$ and $\Sigma_v$ are deterministic but $d_t$ and $F_t$ are time-variant. 

\section*{II. Polynomial diffusion model}

In this section, we present a general framework of the polynomial diffusion model. A simulation study is given in \cite{he2024multi}.

Under the polynomial diffusion framework, the spot price $S_t$ is expressed as a polynomial function of the latent state vector $x_t$ (with components $\chi_t$ and $\xi_t$):
$$S_t = p_n(x_t) = \alpha_1 + \alpha_2 \chi_t + \alpha_3 \xi_t + \alpha_4 \chi_t^2 + \alpha_5 \chi_t \xi_t + \alpha_6 \xi_t^2. $$
In this context, we assume that the polynomial $p_n(x_t)$ has a degree of 2 (or 1 if $\alpha_4 = \alpha_5 = \alpha_6 = 0$). However, it is worth noting that all the theorems presented here are applicable even for polynomials of higher degrees. Additionally, similar to the Schwartz-Smith model, we assume that $\chi_t$ and $\xi_t$ follow an Ornstein–Uhlenbeck process
$$d\chi_t = - \kappa \chi_t dt + \sigma_{\chi} dW_t^{\chi}$$
$$d\xi_t = (\mu_{\xi} - \gamma \xi_t) dt + \sigma_{\xi} dW_t^{\xi}$$
for real processes and
$$d\chi_t = (- \kappa \chi_t - \lambda_{\chi}) dt + \sigma_{\chi} dW_t^{\chi*}$$
$$d\xi_t = (\mu_{\xi} - \gamma \xi_t - \lambda_{\xi}) dt + \sigma_{\xi} dW_t^{\xi*}$$
for risk-neutral processes.

Now, consider any processes that follow the stochastic differential equation
$$dX_t = b(X_t)dt + \sigma(X_t)dW_t,$$
where $W_t$ is a $d$-dimensional standard Brownian Motion and map $\sigma: \mathbb{R}^d \to \mathbb{R}^{d \times d}$ is continuous. Define $a := \sigma \sigma^\top$. For maps $a: \mathbb{R}^d \to \mathbb{S}^{d}$ and $b: \mathbb{R}^d \to \mathbb{R}^d$, suppose we have $a_{ij} \in Pol_2$ and $b_i \in Pol_1$. $\mathbb{S}^d$ is the set of all real symmetric $d \times d$ matrices and $Pol_n$ is the set of all polynomials of degree at most $n$. Then the solution of the SDE is a polynomial diffusion. Moreover, we define the generator $\mathcal{G}$ associated to the polynomial diffusion $X_t$ as
$$\mathcal{G}f(x) = \frac{1}{2} Tr\left( a(x) \nabla^2 f(x)\right) +
b(x)^\top \nabla f(x)$$
for $x \in \mathbb{R}^d$ and any $C^2$ function $f$. $\nabla^2$ and $\nabla$ denote the Hessian and Jacobian matrix of the function $f(\cdot)$, respectively, and $Tr(\cdot)$ denotes the trace of a matrix. Let $N$ be the dimension of $Pol_n$, and $H: \mathbb{R}^d \to \mathbb{R}^N$ be a function whose components form a basis of $Pol_n$. Then for any $p \in Pol_n$, there exists a unique vector $\vec{p} \in \mathbb{R}^N$ such that
$$p(x) = H(x)^\top \vec{p}$$
and $\vec{p}$ is the coordinate representation of $p(x)$. Moreover, there exists a unique matrix representation $G \in \mathbb{R}^{N \times N}$ of the generator $\mathcal{G}$, such that $G \vec{p}$ is the coordinate vector of $\mathcal{G} p$. So we have
$$\mathcal{G} p(x) = H(x)^\top G \vec{p}.$$

\textbf{Theorem 1}: Let $p(x) \in Pol_n$ be a polynomial with coordinate representation $\vec{p} \in \mathbb{R}^N$, $G \in \mathbb{R}^{N \times N}$ be a matrix representation of generator $\mathcal{G}$, and $X_t \in \mathbb{R}^d$ satisfies the SDE. Then for $0 \le t \le T$, we have
$$\mathbb{E} \left[ p(X_T) | \mathcal{F}_t \right] = H(X_t)^\top e^{(T-t)G} \vec{p},$$
where $\mathcal{F}_t$ represents all information available until time $t$.

Obviously, the latent state vector $x_t$ satisfies the SDE with
$$
b(x_t) = \left[ \begin{matrix}
-\kappa \chi_t - \lambda_{\chi} \\
\mu_{\xi} - \gamma \xi_t - \lambda_{\xi}
\end{matrix} \right],
\sigma(x_t) = \left[ \begin{matrix}
\sigma_{\chi} & 0 \\
0 & \sigma_{\xi}
\end{matrix} \right],
a(x_t) = \sigma(x_t) \sigma(x_t)^\top = \left[ \begin{matrix}
\sigma_{\chi}^2 & 0 \\
0 & \sigma_{\xi}^2
\end{matrix} \right].
$$
The basis
$$H(x_t) = (1, \chi_t, \xi_t, \chi_t^2, \chi_t \xi_t, \xi_t^2)^\top$$
has a dimension $N = 6$. The coordinate representation is
$$\vec{p} = (\alpha_1, \alpha_2, \alpha_3, \alpha_4, \alpha_5, \alpha_6)^\top.$$
By applying $\mathcal{G}$ to each element of $H(x_t)$, we get the matrix representation
$$
G = \left[ \begin{matrix}
0 & -\lambda_{\chi} & \mu_{\xi} - \lambda_{\xi} &
\sigma_{\chi}^2 & 0 & \sigma_{\xi}^2 \\
0 & -\kappa & 0 & -2 \lambda_{\chi} & \mu_{\xi} - \lambda_{\xi} & 0 \\
0 & 0 & -\gamma & 0 & -\lambda_{\chi} & 2\mu_{\xi} - 2\lambda_{\xi} \\
0 & 0 & 0 & -2\kappa & 0 & 0 \\
0 & 0 & 0 & 0 & -\kappa - \gamma & 0 \\
0 & 0 & 0 & 0 & 0 & -2\gamma
\end{matrix} \right].
$$
Then, by Theorem 1, the futures price $F_{t,T}$ is given by
$$F_{t,T} = \mathbb{E}^*(S_T | \mathcal{F}_t) = H(x_t)^\top e^{(T-t)G} \vec{p}.$$
Therefore, we have the non-linear state-space model
$$x_t = c + E x_{t-1} + w_t,$$
$$y_t = H(x_t)^\top e^{(T-t)G} \vec{p} + v_t.$$

\section{III. Filtering methods}

\begin{figure}[h]
    \centering
    \includegraphics[width=\textwidth]{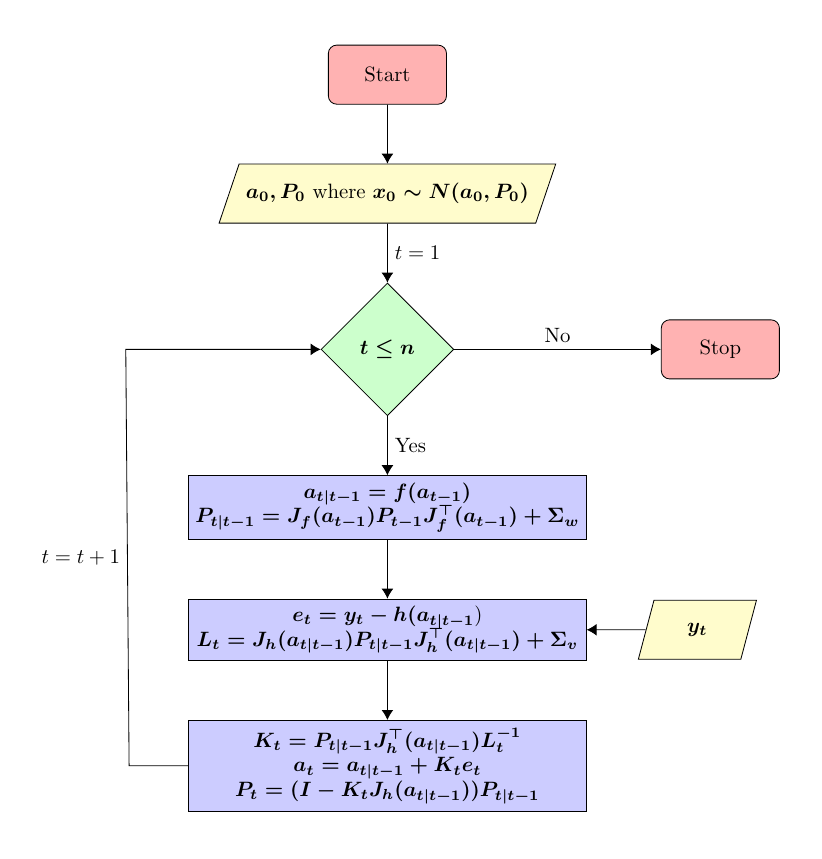}
    \caption{Flowchat of EKF.}
    \label{fig:EKF}
\end{figure}

\begin{figure}[h]
    \centering
    \includegraphics[width=\textwidth]{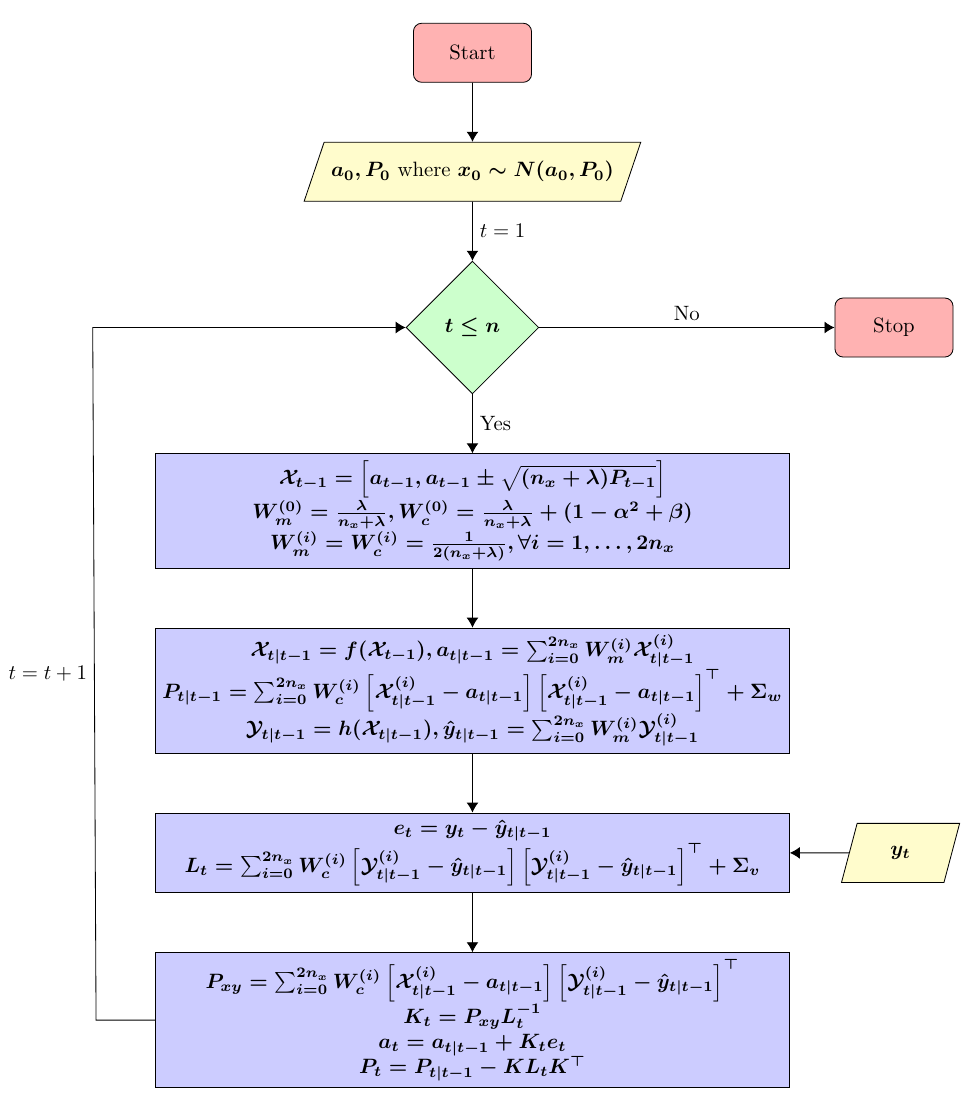}
    \caption{Flowchat of UKF.}
    \label{fig:UKF}
\end{figure}

The Kalman Filter (KF) \cite{harvey1990forecasting} is a commonly used filtering method in estimating latent state variables. However, KF can only deal with the linear Gaussian state model. To capture the non-linear dynamics in the polynomial diffusion model, we use Extended Kalman Filter (EKF) \cite{julier1997new} and Unscented Kalman Filter (UKF) \cite{julier2004unscented,wan2000unscented}. Suppose we have a non-linear state-space model 
$$x_t = f(x_{t-1}) + w_t, w_t \sim N(\textbf{0}, \Sigma_w), $$
$$y_t = h(x_t) + v_t, v_t \sim N(\textbf{0}, \Sigma_v). $$
The EKF linearises the state and measurement equations through the first-order Taylor series. To run KF, we replace $J_f$ and $J_h$ with $E$ and $F_t$ respectively, where $J_f$ and $J_h$ are the Jacobians of $f(\cdot)$ and $h(\cdot)$. In contrast, the UKF uses a set of carefully chosen points, called sigma points, to represent the true distributions of state variables. Then, these sigma points are propagated through the state equation. The flowcharts of EKF and UKF are given in \autoref{fig:EKF} and \autoref{fig:UKF}, where some notations are defined as follow: 
\begin{align}
a_{t|t-1} &:= \mathbb{E}(x_t | \mathcal{F}_{t-1}),& P_{t|t-1} &:= Cov(x_t | \mathcal{F}_{t-1}), \nonumber \\
a_t &:= \mathbb{E}(x_t | \mathcal{F}_t),& P_t &:= Cov(x_t | \mathcal{F}_t). \nonumber
\end{align}
In this application, we use KF for the Schwartz-Smith model, and EKF/UKF for the polynomial diffusion model. 

\FloatBarrier

\section*{IV. Comparison with existing libraries}

The R package ``NFCP'' \cite{aspinall2021nfcp} was developed for multi-factor pricing of commodity futures, which is a generalisation of the Schwartz-Smith model. However, this package doesn't accommodate the polynomial diffusion model. There are no R packages available for PD models currently. 

There are many packages in R for Kalman Filter, for example, ``dse'', ``FKF'', ``sspir'', ``dlm'', ``KFAS'': ``dse'' can only take time-invariant state and measurement transition matrices; ``FKF'' emphasizes computation speed but cannot run smoother; ``sspir'', ``dlm'' and ``KFAS'' have no deterministic inputs in state and measurement equations. For the non-linear state-space model, the functions ``ukf'' and ``ekf'' in package ``bssm'' run the EKF and UKF respectively. However, this package was designed for Bayesian inference where a prior distribution of unknown parameters is required. To achieve the best collaboration of filters and models, we developed functions of KF, EKF and UKF within this code. 

\section*{Implementation and architecture}

The graphical user interface (GUI) provides an easy way for anyone to use the PDSim package, even without programming knowledge. Simply enter the necessary parameters, and PDSim will simulate the data and generate well-designed interactive visualisations. Currently, PDSim supports data simulation from two models: the Schwartz-Smith two-factor model \cite{schwartz2000short} and the polynomial diffusion model \cite{filipovic2016polynomial}. In this section, we first provide the guide to simulate data using GUI and R script. Then, we give an application of fitting WTI crude oil futures data using PDSim.

\section*{I. Implementation of the Schwartz-Smith model}

\begin{figure}[h]
    \centering
    \includegraphics[width=0.6\textwidth]{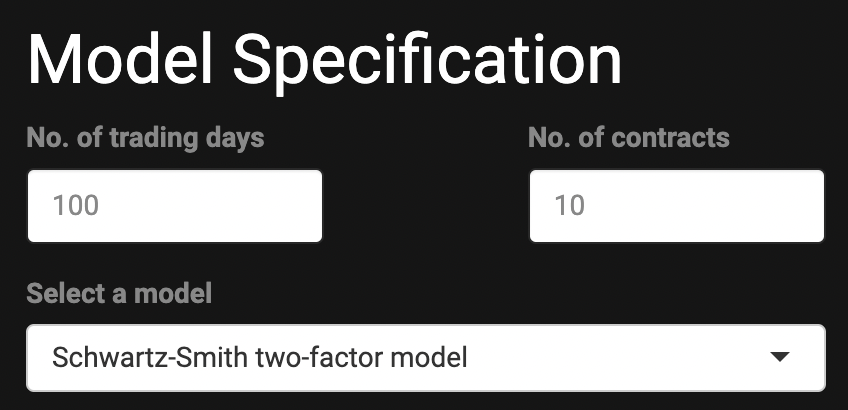}
    \caption{Establish global configurations. }
    \label{fig:SS1}
\end{figure}

First, we establish some global configurations, as illustrated in Figure \ref{fig:SS1}, including defining the number of observations (trading days) and contracts. Additionally, we select the model from which the simulated data will be generated.

\begin{figure}[h]
    \centering
    \includegraphics[width=0.6\textwidth]{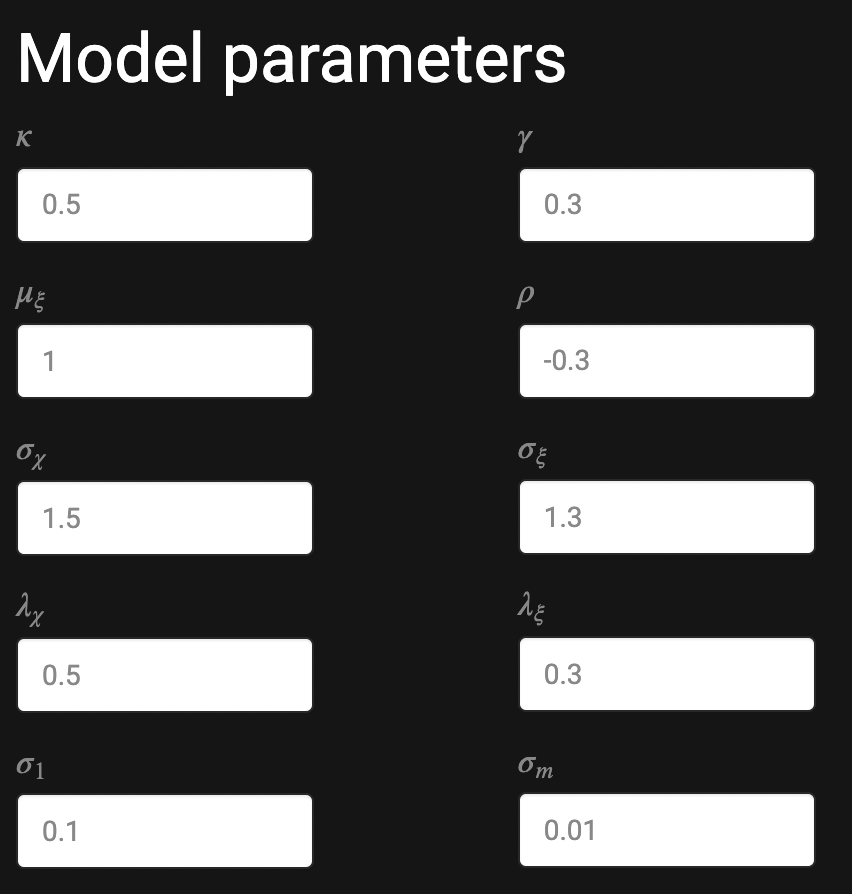}
    \caption{Specify model parameters (Schwartz-Smith model).}
    \label{fig:SS2}
\end{figure}

For the Schwartz-Smith model, we assume that the logarithm of the spot price $S_t$ is the sum of two latent factors:
$$\log{(S_t)} = \chi_t + \xi_t, $$
where $\chi_t$ represents the short term fluctuation and $\xi_t$ is the long term equilibrium price level. Both $\chi_t$ and $\xi_t$ are assumed to follow a risk-neutral mean-reverting process:
$$d\chi_t = (- \kappa \chi_t - \lambda_{\chi}) dt +
\sigma_{\chi} dW_t^{\chi*}, $$
$$d\xi_t = (\mu_{\xi} - \gamma \xi_t - \lambda_{\xi}) dt +
\sigma_{\xi} dW_t^{\xi*}. $$
Here, $\kappa, \gamma \in \mathbb{R}^+$ are the speed of mean-reversion parameters, controlling how quickly these two latent factors converge to their mean levels. Most experiments suggest that $\kappa$ and $\gamma$ fall within the range $(0, 3]$. To avoid issues with parameter identification, we recommend setting $\kappa$ greater than $\gamma$, meaning the short-term fluctuation factor converges faster than the long-term factor. The parameter $\mu_{\xi} \in \mathbb{R}$ represents the mean level of the long-term factor $\xi_t$, and we assume the short-term factor converges to zero. The parameters $\sigma_{\chi}, \sigma_{\xi} \in \mathbb{R}^+$ are volatilities, representing the degree of variation in the price series over time. The parameters $\lambda_{\chi}, \lambda_{\xi} \in \mathbb{R}$ represent risk premiums. While we price commodities under the arbitrage-free assumption, in reality, mean term corrections, represented by $\lambda_{\chi}$ and $\lambda_{\xi}$, are necessary. $W_t^{\chi*}$ and $W_t^{\xi*}$ are correlated standard Brownian Motions with a correlation coefficient $\rho$. In discrete time, these processes are discretised to Gaussian noises, and all parameters are to be specified. Additionally, for simplicity, we assume that the standard errors $\sigma_i, i = 1, \dots, m$ for futures contracts are evenly spaced, i.e., $\sigma_1 - \sigma_2 = \sigma_2 - \sigma_3 = \dots = \sigma_{m-1} - \sigma_m$. All required parameters are shown in Figure \ref{fig:SS2}. 

\begin{figure}[h]
    \centering
    \includegraphics[width=0.6\textwidth]{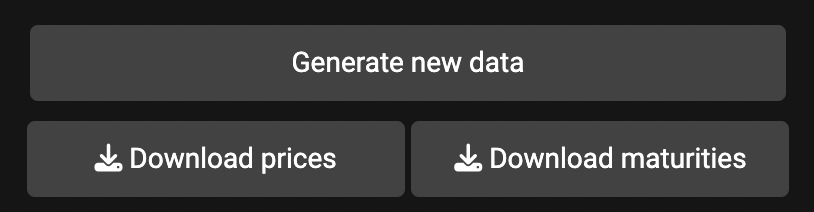}
    \caption{Download simulated data.}
    \label{fig:SS3}
\end{figure}

Finally, all the simulated data can be downloaded. Click the ``Download prices" and ``Download maturities" buttons to download the data for the futures prices and maturities, respectively. Note that although the Schwartz-Smith model specifies the logarithm of the spot price, all downloaded or plotted data are in real prices - they have been exponentiated. The ``Generate new data" button allows users to simulate multiple realisations from the same set of parameters. Clicking on it will generate another set of random noises, resulting in different futures prices. This button is optional if users only need one realisation.Any change in parameters triggers an automatic data update. These three buttons are shown in Figure \ref{fig:SS3}.

\FloatBarrier

\section*{II. Implementation of the polynomial diffusion model}

The procedure for data simulation from the polynomial diffusion model closely mirrors that of the Schwartz-Smith model, but it requires the specification of additional parameters.

\begin{figure}[h]
    \centering
    \includegraphics[width=0.6\textwidth]{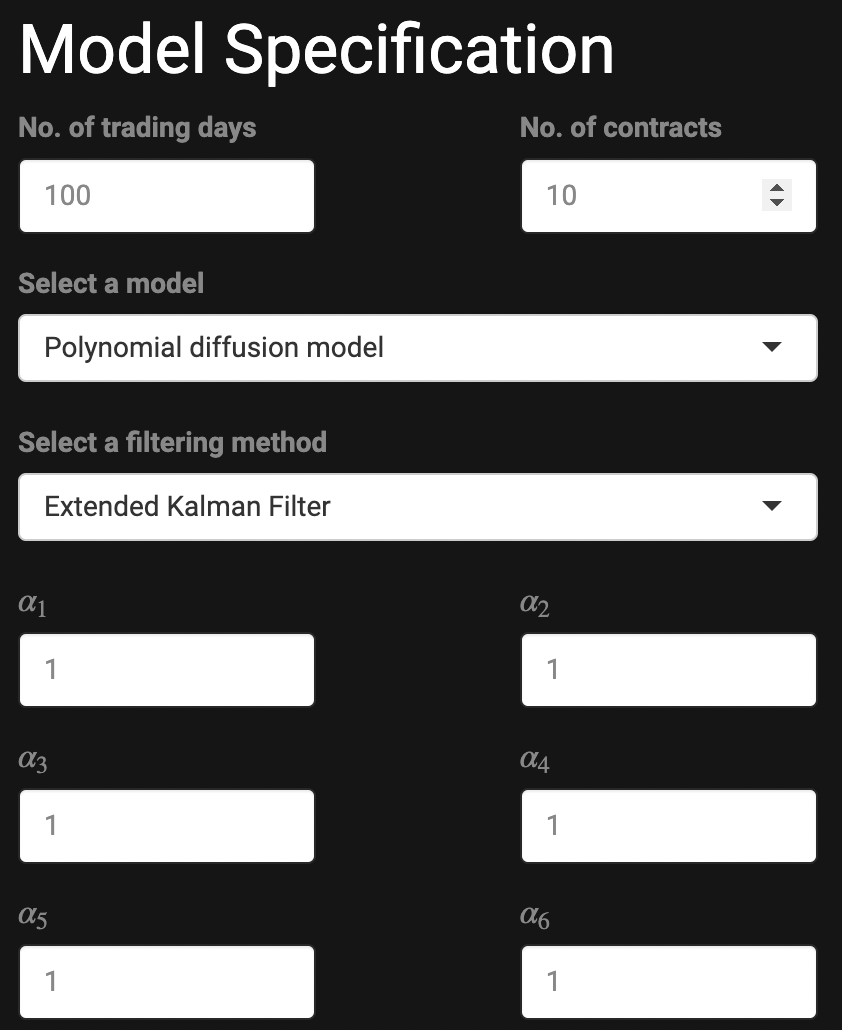}
    \caption{Specify model parameters (polynomial diffusion model). }
    \label{fig:PD1}
\end{figure}

Both the polynomial diffusion model and the Schwartz-Smith model assume that the spot price $S_t$ is the function of two latent factors, $\chi_t$ and $\xi_t$. However, while the Schwartz-Smith model assumes that the logarithm of $S_t$ is the sum of these two factors, the polynomial diffusion model assumes that $S_t$ takes on a polynomial form. Currently, the PDSim GUI supports polynomials of degree 2, which can be expressed as:
$$S_t = \alpha_1 + \alpha_2 \chi_t + \alpha_3 \xi_t + \alpha_4 \chi_t^2 + \alpha_5 \chi_t \xi_t + \alpha_6 \xi_t^2.$$
The coefficients $\alpha_i, i = 1, \dots, 6$ are additional parameters specific to the polynomial diffusion model. If users wish to specify a polynomial of degree 1, they can simply set $\alpha_4 = \alpha_5 = \alpha_6 = 0$. 
Users are also required to choose one of the non-linear filtering methods: the Extended Kalman Filter (EKF) or the Unscented Kalman Filter (UKF). All other procedures for data simulation follow the same steps as those in the Schwartz-Smith model. Additional parameters of the polynomial diffusion model are shown in Figure \ref{fig:PD1}. 

\FloatBarrier

\section*{III. Implementation of the models using R script}

The Shiny application simplifies the simulation and estimation procedures. However, if users require greater control over the simulated data, the R script can be used. This section demonstrates how to use scripts to simulate data through the polynomial diffusion model and estimate the latent state variables.

First, load PDSim and specify the global settings:

\begin{lstlisting}
n_obs <- 100 # number of observations
n_contract <- 10 # number of contracts
dt <- 1/360  # interval between two consecutive time points,
# where 1/360 represents daily data
\end{lstlisting}

Next, specify the parameters and model coefficients:

\begin{lstlisting}
# set of parameters
par <- c(0.5, 0.3, 1, 1.5, 1.3, -0.3, 0.5, 0.3,
         seq(from = 0.1, to = 0.01, length.out = n_contract)) 
x0 <- c(0, 1/0.3) # initial values of state variables
n_coe <- 6 # number of model coefficient
par_coe <- c(1, 1, 1, 1, 1, 1) # model coefficients
\end{lstlisting}

Currently, PDSim supports a polynomial of order 2, which corresponds to six model coefficients. The next step is to define the measurement and state equations. We recommend using ``state\_linear'' and ``measurement\_polynomial'' within the package for a linear state equation and a polynomial measurement equation, respectively. However, users may also define their own functions:

\begin{lstlisting}
# state equation
func_f <- function(xt, par) state_linear(xt, par, dt)
# measurement equation
func_g <- function(xt, par, mats) 
    measurement_polynomial(xt, par, mats, 2, n_coe)
\end{lstlisting}

Finally, simulate the data:

\begin{lstlisting}
dat <- simulate_data(c(par, par_coe), x0, n_obs, n_contract,
                     func_f, func_g, n_coe, "Gaussian", 1234)
price <- dat$yt # simulated futures price
mats <- dat$mats # time to maturity
xt <- dat$xt # state variables
\end{lstlisting}

The latent state variables can then be estimated using either EKF or UKF:

\begin{lstlisting}
est_EKF <- EKF(c(par, par_coe, x0), price, mats, 
               func_f, func_g, dt, n_coe, "Gaussian")
est_UKF <- UKF(c(par, par_coe, x0), price, mats, 
               func_f, func_g, dt, n_coe, "Gaussian")
\end{lstlisting}

\section*{IV. Application to WTI crude oil futures data}

We now demonstrate the implementation of PDSim on real-world data. Specifically, we use weekly WTI crude oil futures prices from 2 January 1990 to 14 February 1995, which is the exact same dataset used in \cite{schwartz2000short}. This dataset is available in the R package ``NFCP''.

For this implementation, we use the following code:

\begin{lstlisting}
library(PDSim)
library(NFCP)
data(SS_oil)
contracts <- SS_oil$stitched_futures
n_obs <- dim(contracts)[1]
n_contract <- dim(contracts)[2]
maturities <- matrix(rep(SS_oil$stitched_TTM, n_obs), 
                     nrow = n_obs, byrow = TRUE)
dt <- SS_oil$dt

################################
##### Schwartz Smith model #####
################################
# state equation
func_f <- function(xt, par) state_linear(xt, par, dt)
# measurement equation
func_g <- function(xt, par, mats) measurement_linear(xt, par, mats)

# initial guesses
par0 <- c(1.5, 1.3, 2, 1.5, 1.5, 0.5, 1, 1, rep(0.1, n_contract))
x0 <- c(0, 0)

# negative log-likelihood
nll <- function(par) KF(par = par, yt = as.matrix(log(contracts)), 
                        mats = maturities, delivery_time = 0, 
                        dt = dt, smoothing = FALSE, 
                        seasonality = "None")$nll

# bounds
lb <- c(1e-04, 1e-04, -5, 1e-04, 1e-04, -0.9999, -5, -5, 
        rep(1e-04, n_contract), -5, -5) # lower bounds
ub <- c(    3,     3,  5,     3,     3,  0.9999,  5,  5, 
        rep(    1, n_contract),  5,  5) # upper bounds 

# parameter estimation
result_SS <- optim(par = c(par0, x0), 
                fn = nll, method = "L-BFGS-B", 
                lower = lb, upper = ub)

# estimated futures contracts
est_SS <- KF(par = result_SS$par, yt = as.matrix(log(contracts)), 
             mats = maturities, delivery_time = 0, dt = dt,
             smoothing = FALSE, seasonality = "None")
yt_hat_SS <- data.frame(exp(func_g(t(est_SS$xt_filter), 
                                   result_SS$par, 
                                   maturities)$y))

# mean absolute error of SS model
mae_SS <- colMeans(abs(log(contracts) - log(yt_hat_SS)))

######################################
##### Polynomial diffusion model #####
######################################
n_coe <- 6

# state equation
func_f <- function(xt, par) state_linear(xt, par, dt)
# measurement equation
func_g <- function(xt, par, mats) 
  measurement_polynomial(xt, par, mats, 2, n_coe)

# initial guesses
par0 <- c(1.5, 0.3, 0, 1.5, 1.5, 0, 0.5, 0.5, rep(0.1, n_contract))
par_coe0 <- c(1, 1, 1, 1, 1, 1)
x0 <- c(0, 0)

# negative log-likelihood
nll_EKF <- function(par) EKF(par = par, yt = as.matrix(contracts), 
                             mats = maturities, func_f = func_f, 
                             func_g = func_g, dt = dt, n_coe = n_coe, 
                             noise = "Gaussian")$nll

# bounds
lb <- c(1e-04, 1e-04, -5, 1e-04, 1e-04, -0.9999, -5, -5, 
        rep(1e-04, n_contract), rep(-5, n_coe), -5, -5) # lower bounds
ub <- c(    3,     3,  5,     3,     3,  0.9999,  5,  5, 
        rep(    1, n_contract), rep( 5, n_coe),  5,  5) # upper bounds 

# parameter estimation
result_PD_EKF <- optim(par = c(par0, par_coe0, x0), 
                   fn = nll_EKF, method = "L-BFGS-B", 
                   lower = lb, upper = ub, 
                   control = list(maxit = 10000))

# estimated futures contracts
est_EKF <- EKF(result_PD_EKF$par, as.matrix(contracts),
               maturities, func_f, func_g, dt, n_coe, "Gaussian")
yt_hat_EKF <- data.frame(func_g(t(est_EKF$xt_filter), 
                                result_PD_EKF$par, maturities)$y)

# mean absolute error of PD model
mae_EKF <- colMeans(abs(log(contracts) - log(yt_hat_EKF))) 
\end{lstlisting}

Table \ref{tbl:mae} reports the mean absolute errors (MAEs) of the logarithm of futures prices estimated using both the Schwartz–Smith (SS) model and the polynomial diffusion (PD) model. For comparison, we also include the results from Schwartz and Smith’s original paper \cite{schwartz2000short}. The MAEs obtained using PDSim for the SS model are comparable to those reported in the original study.

\begin{table}[ht]
    \caption{Estimation mean absolute errors of the logarithm of futures prices with different maturities.}
    \label{tbl:mae}
    \centering
    \begin{tabular}{cccc}
    \toprule
    Maturity & SS paper & PDSim - SS model & PDSim - PD model \\
    \midrule
    1 month & 0.0314 & 0.0268  & 0.0519 \\
    5 months & 0.0035 & 0.0005 & 0.0279 \\
    9 months & 0.0020 & 0.0030 & 0.0290 \\
    13 months & 0.0000 & 0.0000 & 0.0338 \\
    17 months & 0.0028 & 0.0038 & 0.0410 \\
    \bottomrule
    \end{tabular}
\end{table}

The performance of the polynomial diffusion model varies across markets and time periods. In the earlier example (Table \ref{tbl:mae}), the Schwartz–Smith model provides more accurate estimates than the polynomial diffusion model for all contracts. However, Table \ref{tbl:rmse} presents another application of PDSim to WTI crude oil futures data for the period 2015–2018. In this case, the polynomial diffusion model outperforms the Schwartz–Smith model, achieving a mean RMSE of 0.1921.

\begin{table}
\caption{Root mean square errors (RMSE) for each futures contract using the polynomial diffusion (PD) model and the Schwartz-Smith (SS) model for the period from 2015 to 2018.}
\label{tbl:rmse}  
\centering
\begin{tabular}{ccc}
\toprule
Contracts & PDSim - PD model & PDSim - SS model \\
\midrule
Contract 1 & 1.1481 & 0.8884 \\
Contract 2 & 0.7283 & 0.5043 \\
Contract 3 & 0.4262 & 0.2460 \\
Contract 4 & 0.2522 & 0.1189 \\
Contract 5 & 0.1347 & 0.0614 \\
Contract 6 & 0.0516 & 0.0903 \\
Contract 7 & 0.0136 & 0.1056 \\
Contract 8 & 0.0323 & 0.1080 \\
Contract 9 & 0.0385 & 0.0963 \\
Contract 10 & 0.0251 & 0.0708 \\
Contract 11 & 0.0018 & 0.0526 \\
Contract 12 & 0.0322 & 0.0611 \\
Contract 13 & 0.0693 & 0.0933 \\
Mean & 0.2272 & 0.1921 \\
\bottomrule
\end{tabular}
\end{table}

It is worth noting that fitting real data can be computationally expensive. This is primarily due to the high dimensionality of the parameter space, which leads to slow optimisation. For reference, fitting the WTI crude oil futures data using the polynomial diffusion model takes approximately 30 minutes. In this case, we employ the “L-BFGS-B” optimisation method introduced in \cite{byrd1995limited}. We recommend that users experiment with different optimisation algorithms depending on the dataset, in order to achieve more efficient estimation. 

\FloatBarrier

\section*{Quality control}

We conducted several tests to ensure that all functionalities of PDSim are operating correctly.

First, we replicated selected results from \cite{schwartz2000short} using PDSim. The reproduced figures match exactly with the two figures in \cite{schwartz2000short}, as presented in the ``Replicating Schwartz and Smith’s Results'' section on the GitHub page. Moreover, the mean absolute errors estimated by PDSim are very close to those reported in the original paper, as discussed in the ``Application to WTI crude oil futures data'' section of this paper.

Second, we plotted the simulated and estimated futures prices with 95\% confidence intervals for both the Schwartz-Smith model and the polynomial diffusion model. These plots specifically validate the implementation of the filtering methods. Users can reproduce these plots using the code provided in the sections ``Tests for Schwartz-Smith Model" and ``Tests for Polynomial Diffusion Model" on the GitHub page.

Third, users can perform a unit test via the ``Unit Tests" navigation bar of PDSim. This unit test calculates the coverage rate, defined as the proportion of simulated trajectories where over 95\% of points fall within the confidence interval for a specific set of parameters. If the coverage rate exceeds 95\%, we consider the set of parameters to be reasonable. A detailed introduction to the unit test is available in the ``Unit Tests" section on the GitHub page.

Finally, we used the ``tinytest'' package to test the three main functions of PDSim, KF, EKF, and UKF. The outputs from PDSim are compared with corresponding analytical results. Users can re-run these tests after installing PDSim by executing:
\begin{lstlisting}
tinytest::test_package("PDSim")
\end{lstlisting}

\section*{(2) Availability}
\vspace{0.5cm}
\section*{Operating system}

PDSim can be run on any operating system that supports R. Additionally, it can be run on the Shiny server at \url{https://peilunhe.shinyapps.io/pdsim/} or via Docker. 

\section*{Programming language}

R 4.3.0 or above. 

\section*{Additional system requirements}

None.

\section*{Dependencies}

PDSim requires the following R packages: 
\begin{itemize}
    \item DT ($>= 0.31$)
    \item ggplot2 ($>= 3.4.4$)
    \item lubridate ($>= 1.9.3$)
    \item MASS ($>= 7.3.60$)
    \item plotly ($>= 4.10.4$)
    \item scales ($>= 1.3.0$)
    \item shiny ($>= 1.8.0$)
    \item shinythemes ($>= 1.2.0$)
    \item tidyr ($>= 1.3.1$)
\end{itemize}

\section*{List of contributors}

All authors contributed to the software. 

\section*{Software location:}

{\bf Code repository} 
\begin{description}[noitemsep,topsep=0pt]
	\item[Name:] GitHub
	\item[Persistent identifier:] \url{https://github.com/peilun-he/PDSim}
	\item[Licence:] MIT
	\item[Date published:] 14/08/24
\end{description}

\section*{Language}

English

\section*{(3) Reuse potential}

Some of the functions in PDSim were originally developed and used in \cite{he2024multi} to estimate latent state variables and futures prices using the Schwartz-Smith model and the polynomial diffusion model. This package not only allows for detailed estimation but also offers a dynamic web application that vividly visualises the parameter identification challenges discussed in \cite{he2024multi}. It serves as a valuable tool for researchers and practitioners in finance who are focused on commodity futures pricing. The interactive interface provides users with a convenient tool for exploring both models, and analysing their sensitivity to individual parameter changes. The user-friendly design ensures accessibility for all users, even those with no programming experience. Additionally, the downloadable plots can be seamlessly incorporated into professional reports or academic papers.

\section*{(4) Discussions and Limitations}

PDSim consists of both a Shiny application and an R package. The Shiny app provides a straightforward interface for simulating commodity futures data and estimating latent factors, making it accessible to a wide range of users. The R package, in contrast, contains the core functions and offers greater flexibility, allowing users to fit real-world data through R scripts. This separation is primarily due to computational considerations. Parameter estimation in high-dimensional spaces is computationally expensive, making R scripts more suitable than a web-based application for such tasks.

In financial modelling, stochastic processes with jumps are commonly employed. At present, the incorporation of jumps is out of the the scope of the current version of PDSim. However, the inclusion of jump components have been analysed in the following studies, \cite{bernard2008forecasting}, \cite{da2019jump}, or \cite{nguyen2019jumps}, and in particular, the polynomial diffusion model with jumps has been studied in \cite{filipovic2020polynomial} and its practical implementation would significantly broaden the applicability of PDSim. PDSim focuses on adapting and implementing the standard polynomial diffusion model rather than extending it to jump-diffusion settings.

\section*{Acknowledgements}

This study was presented at the 17th International Conference on Computational and Financial Econometrics, Mathematics of Risk 2022, 4th Australasian Commodity Markets Conference, and 24th International Congress on Insurance: Mathematics and Economics. We would like to thank the audiences and organisers for their valuable feedback and suggestions. 

\section*{Funding statement}

This software is not funded by any individuals or institutions. 

\section*{Competing interests}

The authors declare that they have no competing interests.

\catcode`'=9
\catcode``=9
\bibliographystyle{agsm}
\bibpunct{[}{]}{;}{n}{}{,}
\bibliography{references}

\vspace{2cm}

\rule{\textwidth}{1pt}

{ \bf Copyright Notice} \\
Authors who publish with this journal agree to the following terms: \\

Authors retain copyright and grant the journal right of first publication with the work simultaneously licensed under a  \href{http://creativecommons.org/licenses/by/3.0/}{Creative Commons Attribution License} that allows others to share the work with an acknowledgement of the work's authorship and initial publication in this journal. \\

Authors are able to enter into separate, additional contractual arrangements for the non-exclusive distribution of the journal's published version of the work (e.g., post it to an institutional repository or publish it in a book), with an acknowledgement of its initial publication in this journal. \\

By submitting this paper you agree to the terms of this Copyright Notice, which will apply to this submission if and when it is published by this journal.

\end{document}